\documentclass[twocolumn,pre,showpacs]{revtex4}
\usepackage{graphicx}%
\usepackage{psfrag}
\usepackage{dcolumn}
\usepackage{amsmath}
\usepackage{amssymb}
\usepackage{latexsym}
\usepackage{hyperref}
\begin{document}

\def\bea{\begin{eqnarray}}
\def\eea{\end{eqnarray}}
\def\beq{\begin{equation}}
\def\eeq{\end{equation}}
\def\Zp{{Z}_{\parallel}}
\def\Zo{{Z}_{\perp}}
\def\rv{\bf r}
\def\rp{{\bf r}_{\parallel}}
\def\kp{{\bf k}_{\parallel}}
\def\kap{{\vec \kappa}_{\parallel}}
\def\lp{{l}_{\parallel}}
\def\Gp{{G}_{\parallel}}
\def\ro{{r}_{\perp}}
\def\lo{{l}_{\perp}}
\def\Go{{G}_{\perp}}
\def\ioo{I_\perp^0(s,s')}
\def\iot{I_\perp^2(s,s')}
\def\ipo{I_\parallel^0(s,s')}
\def\ipt{I_\parallel^2(s,s')}
\def\Eo{E_\perp (s,s'\mid\ro(0), \ro(L))}
\def\Ep{E_\parallel (s,s'\mid\rp(0), \rp(L))}

\def\f{\frac}
\def\k{\kappa}
\def\sx{\sigma_{xx}}
\def\sy{\sigma_{yy}}
\def\sxy{\sigma_{xy}}
\def\e{\epsilon}
\def\ve{\varepsilon}
\def\ex{\epsilon_{xx}}
\def\ey{\epsilon_{yy}}
\def\exy{\epsilon_{xy}}
\def\be{\beta}
\def\D{\Delta}
\def\h{\theta}
\def\t{\tau}
\def\r{\rho}
\def\a{\alpha}
\def\s{\sigma}
\def\kb{k_B}
\def\la{\langle}
\def\ra{\rangle}
\def\nn{\nonumber}
\def\bu{{\bf u}}
\def\bn{\bar{n}}
\def\br{{\bf r}}
\def\up{\uparrow}
\def\dn{\downarrow}
\def\S{\Sigma}
\def\dg{\dagger}
\def\d{\delta}
\def\p{\partial}
\def\l{\lambda}
\def\G{\Gamma}
\def\o{\omega}
\def\g{\gamma}
\def\kv{\bar{k}}
\def\ha{\hat{A}}
\def\hv{\hat{V}}
\def\hg{\hat{g}}
\def\hG{\hat{G}}
\def\hTT{\hat{T}}
\def\noi{\noindent}
\def\a{\alpha}
\def\d{\delta}
\def\p{\partial} 
\def\r{\rho}
\def\xv{\vec{x}}
\def\qv{\vec{q}}
\def\fv{\vec{f}}
\def\ov{\vec{0}}
\def\vv{\vec{v}}
\def\la{\langle}
\def\ra{\rangle}
\def\e{\epsilon}
\def\o{\omega}
\def\n{\eta}
\def\g{\gamma}
\def\th{\hat{t}}
\def\uh{\hat{u}}
\def\break#1{\pagebreak \vspace*{#1}}
\def\f{\frac}
\def\hf{\frac{1}{2}}
\def\uu{\vec{u}}
\title{On the size and shape of excluded volume polymers confined between parallel plates}
\author{Debasish Chaudhuri}
\email{chaudhuri@amolf.nl} \affiliation{ FOM Institute AMOLF, Science Park 104, 1098XG Amsterdam, The Netherlands
}

\author{Bela Mulder}
\email{mulder@amolf.nl} \affiliation{ FOM Institute AMOLF, Science Park 104, 1098XG Amsterdam, The Netherlands
 }

\date{\today}

\begin{abstract}
A number of recent experiments have provided detailed observations of the
 configurations of long DNA strands under nano-to-micrometer sized confinement.
 We therefore revisit the problem of an excluded volume polymer chain confined
 between two parallel plates with varying plate separation. We show that the
 non-monotonic behavior of the overall size of the chain as a
 function of plate-separation, seen in computer simulations and reproduced
 by earlier theories, can already be predicted on the basis of scaling arguments.
 However, the behavior of the size in a plane parallel to the plates,
 a quantity observed in recent experiments, is predicted to be monotonic,
 in contrast to the experimental findings. We analyze this problem in depth
 with a mean-field approach that maps the confined polymer
 onto an anisotropic Gaussian chain, which allows the size of the polymer
 to be determined separately in the confined and unconfined directions.
 The theory allows the analytical construction of a smooth cross-over between
 the small plate-separation de Gennes regime and the large
 plate-separation Flory regime. The results show good agreement with Langevin
 dynamics simulations, and confirm the scaling predictions.
\end{abstract}
\maketitle
\section{Introduction}

With the typical length scales in cells ranging from tens of
nanometers to tens of microns, their polymeric constituents
are often spatially constrained. Some of the prominent examples are
the cytoskeletal filamentous protein aggregates
microtubules and F-actin, which can have lengths up to tens of microns.
Recent studies on microtubules in plant cells~\cite{Lagomarsino2007}, DNA
packaging in viral capsids~\cite{Fuller2007}, DNA segregation in bacterial cell
division~\cite{Kleckner2004,Bates2005,Jun2006,Liu2009} focused on
properties of biopolymers strongly influenced by confining geometries.
This has naturally led to increased interest in understanding the physics of
strongly confined polymers.
Among all the polymeric constituents of cell, DNA  with its relatively low bending stiffness in comparison to its
extremely long length has a special place as it can under most circumstances be described as an classical excluded
volume polymer~\cite{Gennes1979}. Important examples of confined DNA are (i)~chromosomal DNA that can have bare
lengths up to centimeters trapped inside cell nucleus of size in the range of $1$ -- $10 \mu$m, (ii)~bacterial DNA
of lengths between $0.1$ -- $100 \mu$m trapped within a small bacterial volume which for the case of E.coli is about
$2 \mu$m long and $0.5 \mu$m in diameter, (iii)~mitochondrial DNA of length about $5 \mu$m for human beings with the
available mitochondrial size between $0.5$ -- $10 \mu$m. The combined effects of confinement, self-avoidance and
entropic forces act together in deciding the structure and properties of confined polymers.

A number of single molecule fluorescence microscopy studies have recently
investigated the structure and dynamics of
confined DNA trapped between parallel plates~\cite{Chen2004,Jo2007,Bonthuis2008,Uemura2010} or within cylindrical
geometries~\cite{Tegenfeldt2004, Reisner2005, Stein2006} with the confining dimensions typically of the order of, or
smaller than, the bulk radius of gyration of the polymer.
In particular Bonthuis et al.\ \cite{Bonthuis2008} have
recently observed DNA confined to parallel-plate nanochannels,
showing among others an interesting non-monotonic
variation of the two-dimensional (2d) projected radius of
gyration with the distance between channel walls.

Theoretical analyses of confined excluded volume chains already have a long history, encompassing a number of different
techniques and approaches such as scaling theories~\cite{Daoud1977,Brochard1979}, renormalization group
methods~\cite{Wang1987,Romeis2009}, computer simulations~\cite{Vliet1990, Vliet1992,Chen2004,Jun2008,Jung2009}, and
mean field theory~\cite{Cordeiro1997}. The qualitative picture that emerges from these analyses is simple: With
increased degree of confinement, the polymer size shrinks in the confining direction(s), and expands in the
non-confined direction(s). The expansion in the non-confined directions is due to the excluded volume repulsion
between different polymer segments. According to de Gennes' scaling theory~\cite{Daoud1977}, this expansion
in the strongly confined regime has a
power-law dependence on the length-scale of the confinement. On the other hand, in the limit of very weak or no
confinement the polymer is expected to behave as a free excluded volume polymer approximately obeying the Flory
relationship between polymer length and spatial size in three dimensions (3d)~\cite{Gennes1979}.

Monte-Carlo (MC) simulations of self-avoiding lattice random walks trapped within reflecting walls have
validated the de Gennes scaling predictions~\cite{Vliet1990, Vliet1992}. These simulations also showed a
non-monotonic variation of the mean size of the polymer, measured as the full 3d radius of gyration, as a function
of the inter-wall separation~\cite{Vliet1990}. This non-monotonicity in the radius of gyration was later also
captured by a mean field theory (MFT) calculation~\cite{Cordeiro1997}. However, the experiments reported in Ref.~\cite{Bonthuis2008}
found non-monotonicity in the projected 2d size of the polymer,
rather than its 3d average. To correctly interpret
these results we need a theoretical approach that explicitly allows us to consider the behavior of the polymer in
the confined and non-confined directions separately.

Here we revisit the problem of an excluded volume polymer confined between
two parallel plates with
the latter desideratum in mind. By reviewing the extant scaling approaches, we show that the non-monotonicity of the
3d radius of gyration with plate-separation already follows from a scaling analysis if due care is given to the
details of its application. We then construct a mean-field theory (MFT) for the confined polymer that goes beyond
the existing theory, in that it explicitly takes into account the symmetry breaking induced by the confinement in
the construction of the non-interacting reference polymer. This approach allows the analytical determination of the
prefactors of the scaling relations in the limits of strong and weak confinement respectively. The results of this
approach are then validated by an explicit off-lattice simulation of a confined polymer.
Our main result is that the
size of the polymer transverse to the confining direction, i.e., parallel to the confining planes, grows
\emph{monotonically} with 
decreasing separation between the plates.
The non-monotonicity of the 3d radius of gyration therefore is
 due to the fact that, with decreasing plate-separation, initially
 the polymer-size in the confining direction decreases more rapidly than
 that it grows in the unconfined directions, and eventually the latter becomes dominant.
 

The paper consists of three main sections. We begin with by describing the existing scaling approaches in
Sec.~\ref{scaling}. In Sec.~\ref{mft} we present our variational MFT calculation and its predictions. In
Sec.~\ref{simu} we describe the Langevin dynamics simulation scheme, and compare the results with our MFT
predictions. Finally we conclude in Sec.~\ref{conc} by summarizing our results and providing an outlook. A number of
appendices collect some of the more technical material.

\section{Scaling theories}
\label{scaling}
In this section, we briefly review the existing scaling theories for a polymer confined
between parallel plates. We show that the non-monotonicity of the mean polymer size as a
function of the plate separation, as observed in earlier simulations~\cite{Vliet1990}, is in
fact intrinsic to de Gennes scaling predictions.

The first scaling theory of self-avoiding polymer behavior is due to Flory. Because of their simplicity Flory like
arguments have been extended to other cases, e.g., semiflexible polymers~\cite{Schaefer1980}. Within Flory scaling
theory the free energy of an excluded volume polymer in $d$-dimensions can be written as~\cite{Flory1953},
\beq
\be F \sim \f{R^2}{N l^2} + l^d \f{N^2}{R^d}, \nn
\eeq
where $R$ is the end-to-end distance of the polymer constructed
of $N$ segments, each of size $l$, $\be = 1/\kb T$ with $\kb$ the Boltzmann constant and $T$ the ambient temperature.
The first term describes the entropic elasticity of a free chain, and the second term describes the repulsive
interaction between different segments. Minimizing the free energy with respect to $R$ one obtains the Flory
estimate of the equilibrium polymer size $R \sim N^{3/(d+2)}\, l$. As de Gennes pointed out~\cite{Gennes1979}, the
success of Flory's theory relies on remarkable cancellation of overestimates in both the terms. For polymers confined between parallel plates, a Flory like argument predicts an equilibrium polymer-size in the unconfined directions that agrees
with de Gennes scaling (which we discuss next), but the free energy it predicts is not extensive in
$N$~\cite{Jung2009}.

\begin{figure}[t]
\psfrag{D}{$D$}
\psfrag{R2}{$R^2_{tot}$}
\psfrag{x2    }{$D^2$}
\psfrag{pll    }{$D^{-1/2}$}
\psfrag{combined    }{$D^2+D^{-1/2}$}
\includegraphics[width=8cm]{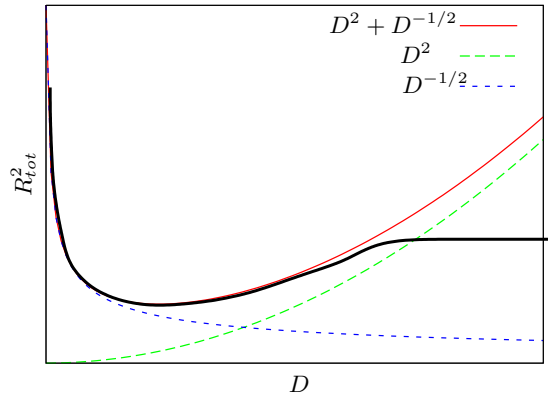}
\caption{(Color online) Schematic diagram of the non-monotonic behavior of average
polymer size $R^2_{tot} \sim a_1 D^2+ a_2 D^{-1/2}$ (dark thick line).
In the limit of small $D ~(\lesssim L$ contour length),
the combination of $D^2$ and $D^{-1/2}$ behavior of the polymer size
in the confining and orthogonal directions, respectively, leads to non-monotonicity
with a clear minimum in the average polymer size $R^2_{tot}$ .
At large $D ~(>L)$,  $R^2_{tot}$ crosses over to the Flory regime and becomes independent of $D$.
}
\label{deGen0}
\end{figure}

The de Gennes scaling arguments for the size of a polymer confined between parallel plates can be described as
follows~\cite{Daoud1977}. If the plate-separation $D \lesssim L=N\,l$ the polymer contour length, the size of the polymer
in the confining direction  (perpendicular to the confining plates),
$R_\perp$, becomes limited by the confinement, so that $R_\perp \sim D$. The size of the
polymer in the unconfined directions (parallel to the confining plates), $R_\parallel$, should be entirely decided by the
following two length scales: The 3d Flory size of a polymer $R \sim N^{3/5}\, l$, and the plate-separation $D$. Thus
$R_\parallel = R \,\phi(R/D)$, so that $\phi(x)=1$ when $x\to 0$ (3d Flory regime) and $\phi(x)\sim x^q$ when $x>1$
(de Gennes regime). In the 2d-limit   ($x\gg 1$) Flory scaling requires $R_\parallel \sim
N^{3/4}\, l$. This ensures $q=1/4$ and therefore
$R_\parallel \sim N^{3/4} (l/D)^{1/4}\, l$. 
This power-law divergence of $R_\parallel$ with decreasing $D$
was observed in MC simulations of self-avoiding
lattice random walks~\cite{Vliet1990,Vliet1992},
and also in recent experiments on confined
DNA~\cite{Bonthuis2008,Uemura2010}.
The excess free energy measured from
the state of an unconfined self-avoiding polymer
can be expressed~\cite{Daoud1977} as $\be F = \psi(R/D)$, so that
$\psi(x)=0$ for $x\to 0$ and $\psi(x)=x^p$ in the other limit of small $D$.
Demanding that the excess free energy has to be an extensive function of $N$ one obtains
$p=5/3$ and $\be F \sim N (l/D)^{5/3}$.

We notice that, in terms of plate-separation $D$ de Gennes scaling predicts two completely  opposite behaviors for
$R_\perp \sim D$ and $R_\parallel \sim N^{3/4} (l/D)^{1/4}\, l$. While $R_\perp$ \emph{decreases}, $R_\parallel$
\emph{increases} with decreasing
$D$. Therefore, 
the mean size
of the polymer, averaged over all the three directions,
$R^2_{tot} = {R_\perp^2 + 2 R_\parallel^2} \sim a_1 D^2 + a_2 D^{-1/2}$
($a_1,\, a_2$ are positive constants) is expected to vary
non-monotonically with $D$.
The $R^2_{tot}\sim D^{-1/2}$ behavior at very small $D$ crosses over to $R^2_{tot}\sim D^2$
at moderately large $D$ via a minimum in $R^2_{tot}$.
This fact is illustrated in Fig.~\ref{deGen0} and was observed
in previous simulations~\cite{Vliet1990}.
At large $D ~(\gg L)$, the scaling function $\phi(R/D)=1$ and
therefore the average size $R^2_{tot}$ becomes independent
of $D$ (Fig.~\ref{deGen0}).
It should be noted here that this behavior is not the same as the
non-monotonicity observed in experimental measurement of
the parallel component of the radius of gyration of a DNA
confined between parallel plates~\cite{Bonthuis2008,Uemura2010}.
de Gennes scaling predicts
a clear monotonic increase of the component $R_\parallel$
with shrinking  $D$.

It is instructive to compare this behavior with that of an
ideal chain confined between parallel plates.
For an ideal chain, in absence of any interaction, the impact of
the confinement remains restricted to the 
confining direction only.
Thus 
$R_\perp \sim D$ for small $D$ and $R_\perp \sim  N^{1/2} l$
in the bulk limit.
However,
$R_\parallel \sim N^{1/2} l$ is obeyed independent of $D$. Therefore
the overall size is $R_{\rm tot}^2 \sim  a_3 D^2 + 2 R_\parallel^2$
with $a_3$  a positive constant.
Starting from the bulk value, 
with shrinking $D$,
$R_{\rm tot}^2$ reduces
to ultimately saturate at $2 R_\parallel^2$.
The initial reduction of overall size 
is thus common to both ideal polymers and excluded volume polymers. However,
the expansion of the overall polymer size with further reduction of $D$
for excluded volume polymers
is a signature of the repulsion between polymer segments.

\section{Mean field theory}
\label{mft}

\subsection{Overview}
The starting point for our MFT is a description of the polymer as a space curve $\mathbf{r}(s)$, in which $s$ is the
arc length parameter with domain $0 \leq s \leq L$. The energetics of the excluded volume (i.e., self-avoiding)
polymer is given by the Edwards Hamiltonian~\cite{Edwards1979}
\bea \beta\mathcal{H} &=&
\frac{3}{2l}\int_{0}^{L}ds\,\left(\f{\p \rv(s)}{\p s}\right)^{2}\crcr
&+& \frac{1}{2}b\int_{0}^{L}ds\int_{0}^{L}ds^{\prime}%
\,\delta\left(\rv(s)-\rv(s^{\prime})\right). \eea The parameter $b$ sets the strength of the repulsive inter-segment
interactions.

Edwards and Singh~\cite{Edwards1979} first proposed to determine the size of a free self-avoiding polymer by mapping
it onto an isotropic Gaussian chain with reference Hamiltonian
\beq
\beta\mathcal{H}_{\mbox{iso}}=\frac{3}{2l_{0}}\int_{0}^{L}ds \left(\f{\p \rv(s)}{\p s}\right)^{2}, \nn
\eeq
and then choosing the effective segment size $l_{0}$
self-consistently by requiring the first order variation of the
mean-squared end-to-end distance $\langle R^2 \rangle$ in the interaction parameter $b$ to vanish. This approach
relatively simply reproduces the Flory result $R \sim N^{\frac{3}{5}}$ for the scaling of polymer size with polymer
length. This same idea was later applied by Thirumalai and co-workers \cite{Cordeiro1997,Morrison2005} to confined
polymers by first imposing the confinement on the isotropic Gaussian reference chain and then performing the
self-consistent calculation of the effective segment length.

One can, however, pose the question whether an isotropic ideal reference chain is appropriate to the situation where
the presence of the boundaries already explicitly breaks the bulk rotational symmetry.
Here we propose to map the
confined polymer onto an \emph{anisotropic} reference chain,
composed of bonds whose length depends on their
absolute orientation with respect to the symmetry axes of the confining geometry. This anisotropic Gaussian reference-chain is described by the Hamiltonian
 \bea
 \beta\mathcal{H}_{0} &=&
\frac{3}{2l_{\parallel}}\int_{0}^{L}ds\, \left(\f{\p \rp(s)}{\p s}\right)^{2}+\frac{3}{2l_{\perp}}\int
_{0}^{L}ds\,\left(\f{\p \ro(s)}{\p s}\right)^{2} \crcr
&\equiv& \beta\mathcal{H}_{\parallel}+\beta\mathcal{H}_{\perp}.
\label{eq:H0}
\eea
The effective segment lengths
$l_{\parallel}$ and $l_{\perp}$ are now the parameters  that have to be determined self-consistently. Apart from the
fact that this choice is physically plausible given the geometry of the system, it has two other potential
advantages. It more explicitly yields independent predictions for the behavior of the polymer in the parallel and
transverse directions with respect to the confinement, and generically multi-parameter mean field theories can be
expected to yield better approximations to the underlying physics than single parameter ones.

We assume the $z$-axis of our coordinate system to be perpendicular to the confining planes. For convenience, we
consider the lateral extent of the system to be finite, defined by a length $W\gg D,$ but freely take the limit of
an infinite extent wherever appropriate.
The hard-wall confinement is implemented through the
Dirichlet boundary condition by which the Green's function
corresponding to the reference Hamiltonian
vanishes at the two confining walls.
The full Hamiltonian can be expressed as
\beq
\beta\mathcal{H}=\beta\mathcal{H}_{0}+\beta\Delta\mathcal{H}_{\parallel}%
+\beta\Delta\mathcal{H}_{\perp}+\beta\Delta\mathcal{H}_{b} \label{eq:Hexp}
\eeq
where%
\begin{align}
\beta\Delta\mathcal{H}_{\parallel}  &  =\frac{3}{2}\left(  \frac{1}{l}%
-\frac{1}{l_{\parallel}}\right)  \int_{0}^{L}ds\,\left(\f{\p \rp(s)}{\p s}\right)^{2},\\
\beta\Delta\mathcal{H}_{\perp}  &  =\frac{3}{2}\left(  \frac{1}{l}-\frac
{1}{l_{\perp}}\right)  \int_{0}^{L}ds\,\left(\f{\p \ro(s)}{\p s}\right)^{2},\\
\beta\Delta\mathcal{H}_{b}  &  =\frac{1}{2}b\int_{0}^{L}ds\int_{0}%
^{L}ds^{\prime}\,\delta\left( \rv(s)-\rv(s^{\prime})\right)
\label{eq:hb}
\end{align}
are treated as perturbative corrections.

Statistical averages with respect to the Hamiltonian $\mathcal{H}$ are
defined through a path integral%
\begin{widetext}
\begin{equation}
\left\langle A\left[  \mathbf{r}(s)\right]  \right\rangle =\frac{1}{Z_{\cal H}}\int
d\mathbf{r}\left(  0\right)  \int d\mathbf{r}\left(  L\right)  \int
_{\mathbf{r}\left(  0\right)  }^{\mathbf{r}\left(  L\right)  }{\cal D}%
[\mathbf{r}(s)]A\left[  \mathbf{r}(s)\right]  \exp\left\{  {-\beta
\mathcal{H}[\mathbf{r}(s)]}\right\}
\end{equation}
\end{widetext}
where the normalization constant is given by the partition function%
\begin{equation}
Z_{\cal H}=\int d\mathbf{r}\left(  0\right)  \int d\mathbf{r}\left(  L\right)
\int_{\mathbf{r}\left(  0\right)  }^{\mathbf{r}\left(  L\right)  }%
{\cal D}[\mathbf{r}(s)]\exp\left\{  {-\beta\mathcal{H}[\mathbf{r}%
(s)]}\right\}.  \label{eq:norm}%
\end{equation}

We now consider the components of the end-to-end separation $\mathbf{R}_{\parallel}=\mathbf{r}_{\mathbf{\parallel
}}(L)-\mathbf{r}_{\mathbf{\parallel}}(0)$ and $R_{\perp}=r_{\perp}\left(
L\right)  -r_{\perp}\left(  0\right)  $ and the averages
$\langle R_{\parallel}^{2}\rangle$ and
$\langle R_{\perp}^{2}\rangle$.
Expanding the Hamiltonian ${\cal H}$ in the above expression
up to linear order in $\D {\cal H}$ around the reference
Hamiltonian ${\cal H}_0$, we obtain
\begin{align}
\left\langle R_{\parallel}^{2}\right\rangle  &  =\left\langle R_{\parallel
}^{2}\right\rangle _{0}-\delta_{\parallel}+\mathcal{O}({\Delta
}{\mathcal{H}}^{2}),\label{eq:Rpar}\\
\left\langle R_{\perp}^{2}\right\rangle  &  =\left\langle R_{\perp}%
^{2}\right\rangle _{0}-\delta_{\perp}+\mathcal{O}({\Delta
}{\mathcal{H}}^{2}) \label{eq:Rprp}%
\end{align}
where
\begin{align}
\delta_{\parallel}  &  =\left\{  \langle R_{\parallel}^{2} \be \Delta
\mathcal{H}_{\parallel}\rangle_{0}-\langle R_{\parallel}^{2}\rangle_{0}%
\langle \be \Delta\mathcal{H}_{\parallel}\rangle_{0}\right\} \label{eq:delpar}\\
&  +\left\{  \langle R_{\parallel}^{2} \be \Delta\mathcal{H}_{\perp}\rangle
_{0}-\langle R_{\parallel}^{2}\rangle_{0}\langle \be \Delta\mathcal{H}_{\perp
}\rangle_{0}\right\} \nonumber\\
&  +\left\{  \langle R_{\parallel}^{2} \be\Delta\mathcal{H}_{b}\rangle_{0}-\langle
R_{\parallel}^{2}\rangle_{0}\langle \be\Delta\mathcal{H}_{b}\rangle_{0}\right\} \nn\\
& \equiv\delta_{\parallel}^{\parallel}+\delta_{\parallel}^{\perp}+\delta
_{\parallel}^{b},\nonumber\\
\delta_{\perp}  &  =\left\{  \langle R_{\perp}^{2} \be\Delta\mathcal{H}%
_{\parallel}\rangle_{0}-\langle R_{\perp}^{2}\rangle_{0}\langle \be\Delta
\mathcal{H}_{\parallel}\rangle_{0}\right\} \label{eq:delprp}\\
&  +\left\{  \langle R_{\perp}^{2} \be \Delta\mathcal{H}_{\perp}\rangle_{0}-\langle
R_{\perp}^{2}\rangle_{0}\langle \be \Delta\mathcal{H}_{\perp}\rangle_{0}\right\}
\nonumber\\
&  +\left\{  \langle R_{\perp}^{2} \be \Delta\mathcal{H}_{b}\rangle_{0}-\langle
R_{\perp}^{2}\rangle_{0}\langle \be \Delta\mathcal{H}_{b}\rangle_{0}\right\} \nn\\
& \equiv\delta_{\perp}^{\parallel}+\delta_{\perp}^{\perp}+\delta_{\perp}%
^{b}.\nonumber
\end{align}
$\la \dots \ra_0$ indicates a statistical average taken with respect to the
reference Hamiltonian ${\cal H}_0$.
In the above expressions, $\d_\parallel^\parallel$, $\d_\parallel^\perp$, $\d_\perp^\perp$ and
$\d_\perp^\parallel$ are the corrections due to the chain anisotropy, and
$\d_\parallel^b$ and $\d_\perp^b$ are the corrections from excluded volume
interactions.
The MFT approximation requires a choice of $\lp$ and $\lo$
so that
$\langle R_{\parallel}^{2}\rangle=\langle R_{\parallel}^{2}\rangle_0$
and
$\langle R_{\perp}^{2}\rangle=\langle R_{\perp}^{2}\rangle_{0}$.
Thus, solving $\delta_{\parallel}=\delta_{\perp}=0$,
one can obtain the
effective segment lengths $l_{\parallel}$ and $l_{\perp}$ in terms of
plate-separation $D$, contour length $L$ and inter-segment interaction strength $b$.

Ref.~\cite{Edwards1979} obtained Flory scaling
within corrections up to linear order in $\be \D \mathcal{H}$. The inclusion
of higher order terms was shown to change the coefficients of scaling form, however,
keeping them stable~\cite{Edwards1979}.
We expect the same to hold in the present problem as well~\cite{Cordeiro1997},
as the nature of interactions used here is exactly the same.

We now briefly summarize the main results we obtained using this MFT.
In the limit of very large plate-separations ($D \to \infty$) the system becomes isotropic with
\beq
\lp = \lo \equiv l_B \sim (bl)^{2/5} L^{1/5}
\label{eq:lplo}
\eeq
and the mean squared end-to-end  separation follows the relation
\beq
\la R^2 \ra = L l_B \sim (b l )^{2/5} L^{6/5}
\eeq
which is independent of $D$ and obeys Flory scaling.

In the limit of strong confinement ($D \to 0$), 
\bea
\lo & \simeq & \f{l}{2}\left(1+\sqrt{1-\f{4bD}{l \lp}} \right) \simeq l
- \f{4\sqrt{2\pi}}{3}\sqrt{\f{b}{l}} \f{D^{3/2}}{L^{1/2}},
\label{eq:lo}\\
\lp & \simeq& \f{l}{2} \left(1 + \sqrt{1 + \f{9 b}{2 \pi l} \f{L}{D}} \right) \simeq \f{3}{2\sqrt{2\pi}} \sqrt{bl} \sqrt{\f{L}{D}}.
\label{eq:lp}
\eea
Using these, we obtain the size of the polymer in the
confining direction
\bea
\la R_\perp^2 \ra \simeq \left(1 - \f{8}{\pi^2} \right) \f{D^2}{2}- \f{D^2}{8} e^{-3{\k} L}
\label{eq:ro}
\eea
with ${\k}=(1/6)\lo (\pi/D)^2$,
and in the unconfined directions
\bea
\la R_\parallel^2 \ra &\simeq&  \f{1}{3} L l  \left(1 + \sqrt{1 + \f{9 b}{2 \pi l} \f{L}{D}} \right)\crcr
&\sim& \sqrt{\f{1}{2\pi}} (bl)^{1/2} L^{3/2} D^{-1/2}.
\label{eq:rp}
\eea
With shrinking plate-separation $D$ the polymer gets compressed in the confining direction and
expands in the free directions.
Up to the leading order (in the small $D$ limit) $\la R_\perp^2\ra \sim D^2$ and
$\la R_\parallel^2 \ra \sim L^{3/2} D^{-1/2}$ obeying de Gennes scaling. Beyond the
leading order, the expressions we found describe a smooth crossover to large $D$
behavior (see Fig.~\ref{deGen1}).

We now present a detailed derivation of the Eqs.~\ref{eq:lplo} -- \ref{eq:rp}.

\subsection{Formulation}

\subsubsection{Green's function}
In order to determine all the above mentioned averages, we require the
Green's function corresponding to the anisotropic reference Hamiltonian $\mathcal{H}_0$ [Eq.~(\ref{eq:H0})] defined as
\begin{equation}
G_0\left(  \mathbf{r,}L|\mathbf{r}^{\prime},0\right)  =\int_{\mathbf{r}^{\prime
}=\mathbf{r}(0)}^{\mathbf{r}=\mathbf{r}(L)}{\cal D}[\mathbf{r}%
(s)]\exp\left\{  {-\beta\mathcal{H}}_{0}{[\mathbf{r}(s)]}\right\}.
\end{equation}
The Green's function obeys the following differential equation \cite{Doi1994},
\beq
\left(\f{\partial}{\partial L} -\f{\lp}{6} \nabla_\parallel^2 -\f{\lo}{6} \f{d^2}{dz^2} \right) G_0 =0.
\label{eq:gf0}
\eeq

Using the Dirichlet boundary condition in the confining direction
$G_0(z=0,D)=0$ and 
free boundary condition in the unconfined directions
(along $x$ and $y$ axes), one obtains a variable
separable form of the Green's function (see \appendixname -\ref{ap-greens})
\beq
G_0= \Gp(\rp,L \mid \rp',0)\,\,\Go(\ro,L \mid \ro',0)
\eeq
where,
\beq
\Gp=\f{3}{2\pi\lp L} e^{-\f{3}{2\lp L} (\rp-\rp')^2},
\eeq
and
\beq
\Go= \f{2}{D}\sum_{n=1}^\infty \sin\left(\f{n\pi}{D} \ro \right) \sin\left(\f{n\pi}{D} \ro' \right)
                                     e^{-{\k} n^2 L},
\eeq
with ${\k}=\f{1}{6}l_\perp \left(\f{\pi}{D}\right)^2$.

\subsubsection{Partition function}
The partition function corresponding to the reference Hamiltonian
${\cal H}_0$ can also be written in a variable-separable form
\beq
Z = \int d\rv(L) \int d\rv(0) ~ G_0(\rv(L), L\mid \rv(0),0) = \Zp \Zo
\eeq
where
\beq
\Zp=\int d\rp \int d\rp' ~ \Gp(\rp, L\mid \rp',0) = W^2
\eeq
and
\bea
\Zo &=& \int d\ro \int d\ro' ~ \Go(\ro, L\mid \ro',0)\crcr
&=&\sum_{n=0}^\infty \f{8D}{(2n+1)^2\pi^2} e^{-{\k} L (2n+1)^2}.
\eea
Remember that $W^2$ denotes the area of the unconfined $xy$-plane, and we 
freely take the limit $W \to \infty$ wherever appropriate.

\subsubsection{End-to-end separation: Contribution from reference Hamiltonian}
Using the above mentioned Green's function $G_0$
one can calculate the end-to-end separations
%
\bea
\langle R_{\parallel}^{2}\rangle_{0}&=&\frac{\int d\mathbf{r}_{\parallel}\int
d\mathbf{r}_{\parallel}^{\prime}\left(  \mathbf{r}_{\parallel}-\mathbf{r}%
_{\parallel}^{\prime}\right)  ^{2}G_{\Vert}\left(  \mathbf{r}_{\Vert
}\mathbf{,}L|\mathbf{r}_{\Vert}^{\prime},0\right)  }{\int d\mathbf{r}%
_{\parallel}\int d\mathbf{r}_{\parallel}^{\prime}G_{\Vert}\left(
\mathbf{r}_{\Vert}\mathbf{,}L|\mathbf{r}_{\Vert}^{\prime},0\right)  } \crcr %
&=&\frac{2}{3}l_{\Vert}L \label{eq:Rpareval}%
\eea
and
\bea
\langle R_{\perp}^{2}\rangle_{0}  & =& \frac{\int_{0}^{D}dr_{\bot}\int_{0}%
^{D}dr_{\bot}^{\prime}\left(  r_{\perp}-r_{\perp}^{\prime\prime}\right)
^{2}G_{\bot}\left(  r_{\perp}\mathbf{,}L|r_{\perp}^{\prime},0\right)  }%
{\int_{0}^{D}dr_{\bot}\int_{0}^{D}dr_{\bot}^{\prime}G_{\bot}\left(  r_{\perp
}\mathbf{,}L|r_{\perp}^{\prime},0\right)  } \crcr
&=& \f{D^2}{2} \f{\nu}{\xi} \label{eq:Rprpeval}
\eea
where,
\bea
{\nu} &= & \sum_{n=0}^{\infty}\frac{1}{\left(  2n+1\right)  ^{2}}\left(  1-\frac{8}{\left(  2n+1\right)
^{2}\pi^{2}}\right)  e^{-{\k} \left(2n+1\right)^{2} L} \crcr
&-& \sum_{n=1}^{\infty}\frac{1}{\left(  2n\right)  ^{2}%
}e^{-{\k} \left(  2n\right)^{2}L},
\label{eq:N}
\eea
and
\beq
{\xi} = \sum_{n=0}^{\infty}\frac{1}{\left(  2n+1\right)  ^{2}}e^{-{\k} \left(  2n+1\right)^{2}L}.
\label{eq:D}
\eeq

\subsubsection{End-to-end separation: Corrections from chain anisotropy}
We now calculate the first order contributions $\delta_{\parallel}^{\parallel
},\delta_{\parallel}^{\perp},\delta_{\perp}^{\parallel}$ and $\delta_{\perp
}^{\perp}$ that are due to the anisotropy of the reference chain.
Using the definition of the mean-squared end-to-end distance %
\begin{widetext}
\beq
\langle R_{\parallel}^{2}\rangle_{0}
=\frac{\int d\mathbf{r}_{\parallel}\int d\mathbf{r}_{\parallel}^{\prime
}\left(  \mathbf{r}_{\parallel}-\mathbf{r}_{\parallel}^{\prime}\right)
^{2}\int_{\mathbf{r}_{\Vert}^{\prime}=\mathbf{r}_{\Vert}(0)}^{\mathbf{r}%
_{\Vert}=\mathbf{r}_{\Vert}(L)}{\cal D}[\mathbf{r}_{\Vert}(s)]\exp\left\{
{-\beta\mathcal{H}}_{\Vert}{[\mathbf{r}(s)]}\right\}  }{\int d\mathbf{r}%
_{\parallel}\int d\mathbf{r}_{\parallel}^{\prime}\int_{\mathbf{r}_{\Vert
}^{\prime}=\mathbf{r}_{\Vert}(0)}^{\mathbf{r}_{\Vert}=\mathbf{r}_{\Vert}%
(L)}\mathcal{D}[\mathbf{r}_{\Vert}(s)]\exp\left\{  {-\beta\mathcal{H}}_{\Vert
}{[\mathbf{r}(s)]}\right\}  } \nn%
\eeq
\end{widetext}
along with the identity
$${\beta\mathcal{H}}_{\Vert}{[\mathbf{r}(s)]=} \alpha_{\Vert}\frac{2}{3}\left(  \frac{1}{l}-\frac
{1}{l_{\parallel}}\right)  ^{-1}{\beta\Delta\mathcal{H}}_{\Vert}%
{[\mathbf{r}(s)]},$$
where $\alpha_{\Vert}=3/2l_\Vert$, 
we obtain
\begin{align}
&-\frac{3}{2}\left(  \frac{1}{l}-\frac{1}{l_{\parallel}}\right)  \frac
{\partial}{\partial\alpha_{\Vert}}\langle R_{\parallel}^{2}\rangle_{0} \nn\\
&=\left\{  \langle R_{\parallel}^{2}\, \be \Delta\mathcal{H}_{\parallel}%
\rangle_{0}-\langle R_{\parallel}^{2}\rangle_{0}\langle \be \Delta\mathcal{H}%
_{\parallel}\rangle_{0}\right\}  \nn\\
&\mbox{or,~~}\delta_{\parallel}^{\parallel} =\frac{2}{3}\left(  \frac{1}{l}-\frac
{1}{l_{\parallel}}\right)  Ll_{\parallel}^{2}.
\label{eq:dpll}
\end{align}
Similarly, one can show that (with $\a_\perp=3/2l_\perp$),
\begin{align}
\delta_{\parallel}^{\perp}  &  =-\frac{3}{2}\left(  \frac{1}{l}-\frac
{1}{l_{\bot}}\right)  \frac{\partial}{\partial\alpha_{\bot}}\langle
R_{\parallel}^{2}\rangle_{0}=0\\
\delta_{\perp}^{\parallel}  &  =-\frac{3}{2}\left(  \frac{1}{l}-\frac
{1}{l_{\parallel}}\right)  \frac{\partial}{\partial\alpha_{\Vert}}\langle
R_{\bot}^{2}\rangle_{0}=0\\
\delta_{\perp}^{\perp}  &  =-\frac{3}{2}\left(  \frac{1}{l}-\frac{1}{l_{\bot}%
}\right)  \frac{\partial}{\partial\alpha_{\bot}}\langle R_{\bot}^{2}%
\rangle_{0}%
\label{eq:doo}
\end{align}
where the last derivative can be explicitly evaluated by considering
Eq.~(\ref{eq:Rprpeval}). This leads to the following expression,
\bea
\d_\perp^\perp &=& -\f{\pi^2}{12 \xi}~ l_\perp^2 L~ \left(\f{1}{l}-\f{1}{l_\perp} \right)~\crcr
&\times& \left[ \sum_{n=0}^\infty \left(1-\f{8}{(2n+1)^2\pi^2} - \f{\nu}{\xi}\right) e^{-{\k} L(2n+1)^2} \right.\crcr
&-& \left. \sum_{n=1}^\infty e^{-{\k} L (2n)^2}\right] \label{eq:dprp}
\eea
where ${\nu}$ and $\xi$ are defined by Eq.~(\ref{eq:N}) and Eq.~(\ref{eq:D}) respectively.

\subsubsection{End-to-end separation:  Corrections from excluded volume interactions}
Now we evaluate $\d_\parallel^b $ and $\d_\perp^b$.
Note that the presence of $\d (\rv(s) - \rv(s'))$ in the self-avoidance interaction
$\be \D {\cal H}_b$ [Eq.~(\ref{eq:hb})] couples
the transverse and the longitudinal modes of the Green's function. This is the term through which
$\la R_\parallel^2\ra$ becomes dependent on the plate-separation $D$.

The calculation of these two terms are lengthy but straightforward. 
Here, we only quote the final results
deferring the details of the calculations to \appendixname-\ref{ap:exvol}:
\bea
\d_\parallel^b &=& -Z^{-1} \f{b}{\pi}~W^2~ \sum_{n_L,n',n_0=1}^\infty M(n_L,n',n_0;{\k}) \nn\\
           &\times& [2\d_{n_0,n_L} + \d_{n',(n_0+nL)/2} - \d_{n',(n_0-nL)/2} \crcr
           &&   -\d_{n',(nL-n_0)/2} ] \times J_0(n_L, n_0)
\label{dpll_b}
\eea
where,
\bea
&& M(n_L,n',n_0;{\k}) = e^{-{\k} L n_L^2}\times\crcr
&& \int_0^L ds' \int_{s'}^L ds
                   \left[e^{-{\k} s (n'^2-n_L^2)} e^{-{\k} s' (n_0^2-n'^2)}\right],
\eea
\beq
J_0(n_L, n_0) =  \f{1}{n_0 n_L\pi^2}  [1-(-1)^{n_L}] [1-(-1)^{n_0}];
\eeq
and
\bea
\d_\perp^b &=& \f{b}{Z} \,W^2\, \f{3}{2\pi\lp} \sum_{n_L,n',n_0=1}^\infty D^2\,  K(n_L,n',n_0;{\k}) \nn\\
           &\times& [2\d_{n_0,n_L} + \d_{n',(n_0+nL)/2} \crcr
            &-& \d_{n',(n_0-nL)/2}  -\d_{n',(nL-n_0)/2} ] \crcr
           &\times&  \left[ J(n_L,n_0) - \f{\la R_\perp^2 \ra_0}{D^2} J_0(n_L,n_0) \right]
\label{dporth}
\eea
where,
\bea
K(n_L,n',n_0;{\k}) &=& e^{-{\k} L n_L^2} \lim_{a\to 0} \int_0^L ds' \int_{s'}^L ds \crcr
                 && \left[\f{e^{-{\k} s (n'^2-n_L^2)} e^{-{\k} s' (n_0^2-n'^2)}}{(s-s')+a}\right],
                 \label{eq:K}
\eea
\bea
J(n_L,n_0) &=& - \f{1}{n_L^3 n_0\pi^4} [(-1)^{n_L} n_L^2\pi^2 +2(1-(-1)^{n_L})]\nn \\
                     &\times& [1-(-1)^{n_0}] - \f{2}{n_L n_0 \pi^2} (-1)^{n_0}(-1)^{n_L}\nn\\
                     &-& \f{1}{n_L n_0^3\pi^4} [(-1)^{n_0} n_0^2\pi^2 +2(1-(-1)^{n_0})] \nn\\
                     &\times& [1-(-1)^{n_L}].
\label{eq:K}
\eea
\subsection{Results}
\subsubsection{Large plate-separation limit: $D\to \infty$}
Here we  demonstrate that the MFT scheme naturally leads to
Flory scaling in the bulk limit of $D\to\infty$.
This limit is better appreciated, in a
shifted coordinate system in which the confining hard walls are at $z=\pm D/2$.
In this coordinate system, the transverse component of Green's function
is,
\bea
\Go &=& \f{1}{D} \sum_{n,~\mbox{even}} \sin \f{n\pi z'}{D} \sin \f{n\pi z}{D} e^{-{\k} n^2 L}\crcr
&+& \f{1}{D} \sum_{n,~\mbox{odd}}  \cos \f{n\pi z'}{D} \cos \f{n\pi z}{D} e^{-{\k} n^2 L}.\nn
\eea
In the continuum limit of $D\to\infty$
\bea
\Go 
= \sqrt{\f{3}{2\pi\lo L}} \exp \left(-\f{3}{2}\f{(z-z')^2}{\lo L}\right).
\eea
%
$\d_\parallel^\parallel$ is independent of $D$ and is given by Eq.~\ref{eq:dpll}.
The details of the calculation of
$\d_\parallel^b$, $\d_\perp^b$ and
$\d_\perp^\perp$ in this bulk limit
are presented in \appendixname-\ref{apfree}.
In this limit
the equations, $\d_\parallel =\d_\parallel^\parallel+\d_\parallel^b=0$
and $\d_\perp = \d_\perp^\perp+\d_\perp^b =0$
imply,
\bea
\f{L\lp^2}{l} \sim \f{b}{{\lo}^{1/2}} L^{3/2} \label{eq:lplo1}
\eea
and
\bea
\f{L \lo^2}{l} \sim \f{b\lo^{1/2}}{\lp} L^{3/2} \label{eq:lplo2}
\eea
respectively.
Eq.~(\ref{eq:lplo1}) and Eq.~(\ref{eq:lplo2}) leads to $\lp=\lo \equiv l_B$,
the isotropy expected in the  limit of $D\to \infty$.
This also implies
\bea
l_B \sim (bl)^{2/5} L^{1/5}.
\eea
and
\beq
\la R^2 \ra = L l_B \sim (bl)^{2/5} L^{6/5},
\eeq
i.e., Flory scaling in three dimensions.

\subsubsection{Narrow plate-separation limit: $D\to 0$}
In the limit of $D\to 0$, ${\k}=\lo (\pi/D)^2/6 \to \infty$ and only the
small $n$ eigenfunctions 
contribute to the calculation of perturbative corrections to
end-to-end distance.
The simplest approximation in this limit is
the ground state approximation (only the minimum value of $n$'s contribute).

{\em Ground state approximation:}
We first calculate $\d_\parallel^b$ within the ground state approximation. Using
$n_0=n'=n_L=1$, we get $M(1,1,1;{\k})= (L^2/2)\exp(-{\k} L)$  and
$Z=  W^2 ({8D}/{\pi^2}) \exp(-{\k} L)$ which leads to
$\d_\parallel^b = -({3b}/{4\pi}) ({L^2}/{D})$ (using Eq.~\ref{dpll_b}).
Within MFT, $\lp$ satisfies the condition
$\d_\parallel = \d_\parallel^\parallel+\d_\parallel^b=0$,
which implies that,
\beq
\f{2}{3}\left(\f{1}{l} -\f{1}{\lp} \right)L \lp^2 = \f{3b}{4\pi} \f{L^2}{D}.\nn
\eeq

The solution of this equation gives
\beq
\lp = \f{l}{2} \left(1 + \sqrt{1 + \f{9 b}{2 \pi l} \f{L}{D}} \right),\nn
\eeq
which is Eq.~\ref{eq:lp}.
Therefore, we find the relation given in Eq.~\ref{eq:rp}
\bea
\la R_\parallel^2 \ra &=& \f{2}{3} \lp L = \f{1}{3} L l  \left(1 + \sqrt{1 + \f{9 b}{2 \pi l} \f{L}{D}} \right)\crcr
&\sim& \sqrt{\f{1}{2\pi}} (bl)^{1/2} L^{3/2} D^{-1/2}. \nn
\eea
In the last step, the diverging part of $\la R_\parallel^2 \ra_0$  in the limit of
$D\to 0$ is extracted.
Thus, up to the leading order in the limit of $D\to 0$ we find de Gennes scaling
$R_\parallel = \sqrt{\la R_\parallel^2 \ra} \sim (1/\sqrt{2\pi})(bl)^{1/4} L^{3/4} D^{-1/4}$.

Now we calculate the  perpendicular component $\la R_\perp^2 \ra$
within the ground state approximation $n_L=n'=n_0=1$.
%
Thus Eq.~\ref{dporth} reduces to 
\bea
\d_\perp^b = \f{9b}{2\pi\lp\, Z} \,W^2\, D^2 K(1,1,1;{\k})
     \left[J(1,1)-\f{4}{\pi^2} \f{\la R_\perp^2\ra_0 }{D^2}\right]. \nn
\eea
In the expression of $\la R_\perp^2\ra_0$ [Eq.~(\ref{eq:Rprpeval})]
we retain only the $n=0$ terms, within the ground state approximation,
 to obtain
$ \la R_\perp^2\ra_0 = 1-8/\pi^2$.
This leads to $J(1,1)-(4/\pi^2)  \la R_\perp^2\ra_0 = 0$ and therefore, $\d_\perp^b=0$.
It can be easily seen that within the ground state approximation $\d_\perp^\perp=0$ too.
Thus $\lo$ remains indeterminate. 
We need to go up to the first excited state approximation (next higher values of $n$),
to obtain the expression for $\lo$.

{\em First excited state approimation:}
Within this approximation, we can write Eq.~\ref{dporth} as
\bea
\d_\perp^b &=& \f{3b}{2\pi \lp\, Z} W^2 D^2 \left[3 K(1,1,1;{\k})\, T_1 +  2 K(1,2,1;{\k})\, T_1 \right.\crcr
&+& \left. 2 K(2,1,2;{\k}) T_2 + 3 K(2,2,2;{\k}) T_2 \right] \nn
\eea
where,
\bea
T_1 &=& J(1,1)-\f{\la R_\perp^2\ra_0}{D^2} J_0(1,1) = \f{1}{2\pi^2} e^{-3{\k} L},\crcr
T_2 &=& J(2,2)-\f{\la R_\perp^2\ra_0}{D^2} J_0(2,2) = -\f{1}{2\pi^2}.\nn
\eea
We find,
\beq
K(1,1,1;{\k}) \simeq  L e^{-{\k} L}
\left[\log\left(\f{L}{a}\right) -1\right].\nn
\eeq
We observe that the leading order behavior of $K(1,1,1;{\k}) \sim L \exp(-{\k} L)$
apart from a weak logarithmic divergence coming from the $\d$-function nature of the
inter-segment repulsion ($a \to 0$). 
Therefore, $K(1,1,1;{\k})T_1 \sim L\exp(-4 {\k} L)$.
Similarly one can show that all the terms in the above expression of $\d_\perp^b$
has the same leading order behavior.
Using $Z=W^2\, (8D/\pi^2)\exp(-{\k} L)$ we find that, 
$\d_\perp^b \sim (b L D/\lp) \exp(-3{\k} L)$.

Within the first excited state approximation, we can write Eq.~\ref{eq:dprp} as
\beq
\d_\perp^\perp = \f{\pi^2}{16} e^{-3{\k} L} \lo^2 L \left(\f{1}{\lo} -\f{1}{l} \right).\nn
\eeq
Thus the mean field condition, $\d_\perp=\d_\perp^\perp+\d_\perp^b =0$ 
implies,
\bea
\lo \simeq  \f{l}{2}\left[1+\sqrt{1-\f{4bD}{l \lp}} \right]\nn
\eea
which is Eq.~\ref{eq:lo}.
In the limit of $D\to 0$ we can write
$\lo \simeq l-2b D/\lp \simeq l - (4\sqrt{2\pi}/3)\sqrt{b/l} D^{3/2} L^{-1/2}$.
Thus the leading order behavior is $\lo \sim l$ with correction vanishing as $D^{3/2}$.
Using this $\lo$ in ${\k}=(1/6)\lo (\pi/D)^2$, we find
\beq
\la R_\perp^2\ra \simeq \f{D^2}{2} \left[ \left( 1- \f{8}{\pi^2}\right)\right] - \f{D^2}{8} e^{-3{\k} L},\nn
\eeq
which is Eq.~\ref{eq:ro}.
The above expression means that,  the leading order behavior (at smallest $D$ values) of
$$\la R_\perp^2\ra = \f{D^2}{2}  \left[ \left( 1- \f{8}{\pi^2}\right)\right]$$
has corrections in the next order (at larger $D$), which has a complicated functional dependence on
inter-segment interaction strength $b$,  plate-separation $D$ and polymer contour-length
$L$.

Before ending this section, we briefly discuss the pure 2d limit
of the above-mentioned calculation.
Variational MFT calculation for
pure 2d leads to the mean squared end-to-end distance
\beq
{\la R_{2d}^2\ra} \sim (b_0 l)^{1/2} L^{3/2},
\eeq
which obeys 2d Flory scaling. In the above expression $b_0$
measures the strength of the inter-segment repulsion and is dimensionless,
in contrast to the same parameter in 3d denoted by $b$ which has
the dimension of length. This indicates that $l$ is the only
intrinsic microscopic length scale in the system.
Requiring $b=b_0 l$, we find that
$\la R_\parallel^2 \ra$ reaches the 2d limit of $\la R_{2d}^2\ra$
when $D \sim l$ (see \appendixname-\ref{ap:2d}).



\subsubsection{Non-monotonicity in overall polymer-size: Crossover}
\label{nonmono}
Clearly, the mean squared end-to-end distance 
$\la R_{\rm tot}^2 \ra = 2\la R_\parallel^2 \ra + \la R_\perp^2 \ra$
should show a minimum as a function of changing plate-separation $D$. This comes about, because of
the completely opposing behaviors of the two components -- while the size in the
confining direction $\la R_\perp^2 \ra$ shrinks with decreasing $D$, the polymer-size in the
unconfined directions $\la R_\parallel^2 \ra$ expand. Keeping up to the leading order behavior in
$\la R_\perp^2 \ra$ [$=(1-8/\pi^2)D^2/2$] one can easily find the cross-over plate-separation (using Eqs.~\ref{eq:ro} and \ref{eq:rp})
\beq
D_c = (2\pi)^{-1/5}\left( 1-\f{8}{\pi^2} \right)^{-2/5} (b l)^{1/5} L^{3/5}
\label{eq:dc}
\eeq
where $\la R_{\rm tot}^2\ra$ reaches the minimum.
With enhanced degree of confinement (reduced $D$) the total size of the
polymer $\la R_\perp^2 \ra$ first shrinks and when $D<D_c$ the size starts to grow.

$D_c$ follows the same 3d Flory-scaling form as the size of an excluded
volume polymer freely floating in 3d. Thus the position of this minimum
scaled by the bulk polymer size (end-to-end distance or radius of gyration)
$D_c/L^{3/5}$ should be independent of polymer contour-length $L$.
The MC simulation results shown in
Fig.1 of Ref.\cite{Vliet1990} corroborate this fact.

\section{Simulations}
\label{simu} In the previous two sections we demonstrated scaling arguments and a mean field theory to obtain
estimates for the size and shape of confined excluded volume polymers. To validate our mean field picture, we
present the results of a full numerical simulation for such a system.

\subsection{Method}
We perform molecular dynamics simulations of a self  avoiding bead-spring chain
trapped between two soft-walls in presence of a Langevin heat bath.
All the beads interact via the fully repulsive part of Lennard-Jonnes
potential
$V_{LJ}^{rep}(r)=4\e \left[ (\s/r)^{12} - (\s/r)^6 +1/4\right]$,
with a cut-off distance set to $r_c = 2^{1/6} \s$ so that $V_{LJ}^{rep}(r\geq r_c) = 0$.
$\e$ is the strength and $\s$ is the range of the interaction.
The bond between two neighboring beads 
are modeled by a {\em shifted} harmonic potential $V_{sp}(r)=(A/2)(r-\s)^2$
with spring constant $A=100 \e/\s^2$
such that the equilibrium bond-length, in absence of any other forces, is $\s$.
$r$ 
denotes the center-to-center distance between a pair of beads.
We assume that the polymer is confined within two parallel walls placed at  $z=-D/2,D/2$.
The repulsive potential due to the walls is assumed to be the integrated
and shifted Lennard-Jones potential
$V_{wall}(\d z)=\e\left[(\s/\d z)^{10} -(\s/\d z)^4 + g\right]$
with $g=(5/2-1)(2/5)^{5/3}$ and a cut-off distance $z_c= (5/2)^{1/6} \s$,
 so that $V_{wall}(|\d z| \geq z_c) = 0$.
Here $\d z$ is the $z$-separation of
a bead from any one of the walls (top or bottom).
$\e$ and $\s$ set the energy and length scales, respectively.
We integrate Langevin equations using the velocity-Verlet algorithm\cite{Frenkel2002}
with a time-step $\d t=0.01 \t$ where $\t = \s\sqrt{m/\e}$ is 
the characteristic time scale. 
We set the mass of each bead $m=1$
and isotropic friction coefficient $\g=1/\t$.
The Langevin thermostat\cite{Grest1986} is used to
keep the system at a constant temperature $T=1.0\,\e/\kb$.

\subsection{Results}
The  quantity we are interested in is the equilibrium size and
shape  of the polymer. In the model we simulate,
the polymer is made of $i=1,\dots,N$ beads with a mean contour length
$L=(N-1)$ in units of $\s$.
We follow the mean squared end-to-end distance
in the confining $z$-direction $\la Z^2 \ra = \la (z_N - z_1)^2 \ra $ as well as in the
unconfined  $xy$-plane  $\la X^2\ra= \la (x_N - x_1)^2\ra$, $\la Y^2 \ra= \la (y_N - y_1)^2\ra$.
The time evolution of these quantities allows us to identify the equilibration of the system.
In order to test the validity of our simulation scheme,
in a separate simulation of the bulk-system
(using periodic boundary conditions in all three directions)
we measured
$\la R_{\rm bulk}^2 \ra = \la X^2\ra + \la Y^2 \ra + \la Z^2 \ra$
as a function of polymer contour length
$L = 8,\, 16,\, 32,\, 64,\, 128,\, 256$  to obtain Flory scaling
$\la R_{\rm bulk}^2 \ra \sim L^{6/5}$ (data not shown).

\begin{figure}[t]
\psfrag{D}{$D~~~~$}
\psfrag{dr2}{$\la R_i^2 \ra$}
\psfrag{dx2    }{\scriptsize{$\la X^2\ra'$}}
\psfrag{dy2    }{\scriptsize{$\la Y^2\ra'$}}
\psfrag{dz2    }{\scriptsize{$\la Z^2\ra'$}}
\psfrag{bx2    }{\scriptsize{$\g D^2$}}
\psfrag{c/x14   }{\scriptsize{$\mu/D^{1/2}$}}
\psfrag{phix   }{\scriptsize{$\chi(D)$}}
\includegraphics[width=8.4cm]{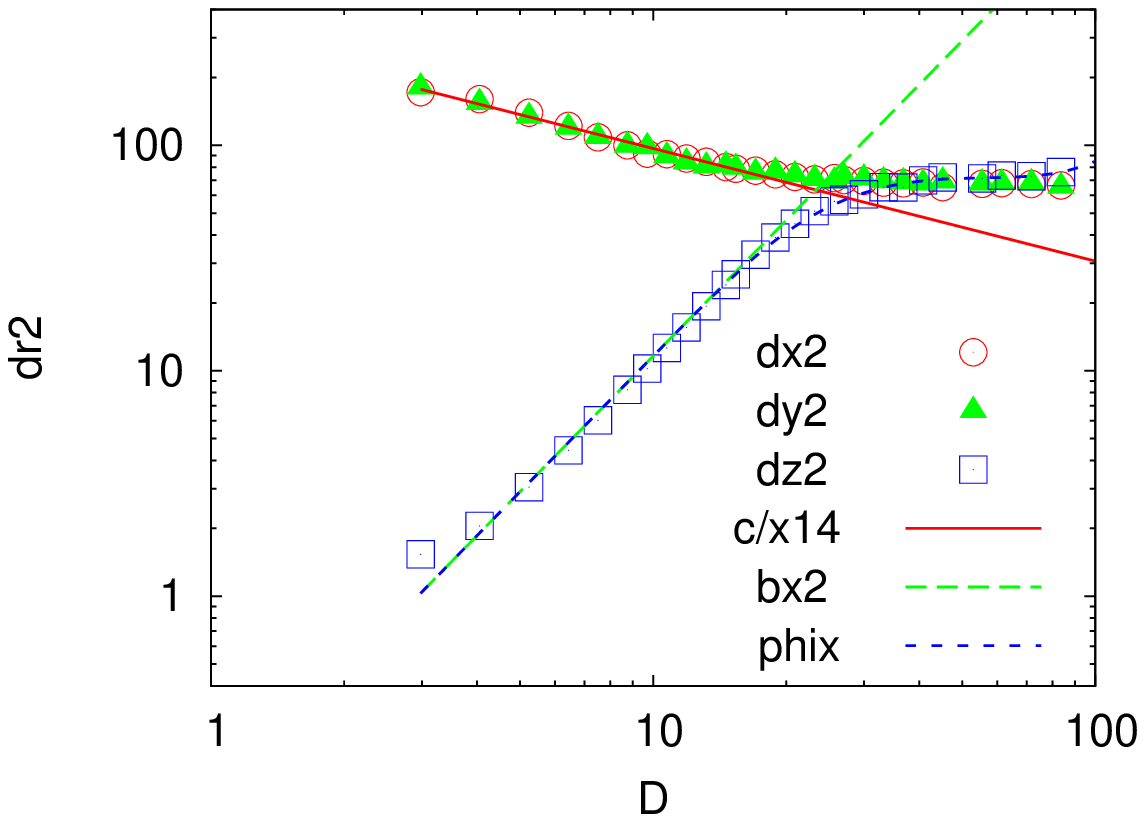} 
\caption{(Color online)
Components of mean squared end-to-end distance  $\la R_i^2\ra$
as a function of plate-separation $D$. All lengths are expressed in units of $\s$.
We have plotted the components $\la X^2\ra'=\la X^2\ra-\l_\parallel$,
$\la Y^2\ra'=\la Y^2\ra-\l_\parallel$, and $\la Z^2\ra'=\la Z^2\ra+\l_\perp$ where
$\l_\parallel$ and $\l_\perp$ are offset polymer sizes (see main text).
With decreasing $D$,
$\la X^2\ra$ and $\la Y^2\ra$ increases in the
same manner while  $\la Z^2\ra$ decreases.
All the components obey de Gennes scaling, up to the leading order:
$\la X^2\ra'=\la Y^2\ra'= \mu/D^{1/2}$, and
$\la Z^2\ra'=\g D^2$ where $\mu=306$ and $\g=0.12$.
The fitted values of offsets are $\l_\parallel=12.8 \pm 4.2$
and $\l_\perp = 1.4$.
At larger $D$,
$\la Z^2\ra'$ shifts from the scaling form (Eq.\ref{eq:ro}) as
$\chi(D)=\g D^2 - c_1 D^2 \exp(-c_2/D^2 + c_3/D^{1/2})$
where  $c_1=0.1$,
$c_2=970$ and $c_3=2.06 \pm 0.36$.
The fitting error in all the parameters is less than $5\%$, unless
specified otherwise.
}
\label{deGen1}
\end{figure}

\begin{figure}[t]
\psfrag{D}{$D$}
\psfrag{<r^2>}{$\la R_{\rm tot}^2 \ra$}
\includegraphics[width=8.4cm]{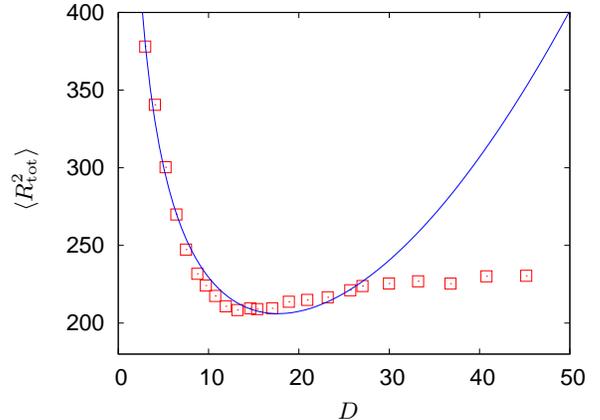}
\caption{(Color online)
Mean squared end-to-end distance $\la R_{\rm tot}^2 \ra = \la X^2\ra + \la Y^2 \ra + \la Z^2 \ra$ as a function of plate-separation $D$. The line is
a plot of $2 (\mu/D^{1/2}+\l_\parallel)+(\g D^2-\l_\perp)$ where we used the
same values of $\mu$, $\g$, $\l_\parallel$ and $\l_\perp$ as in Fig.\ref{deGen1}.
}
\label{deGen2}
\end{figure}

For the simulations of confined system we used an $N=64$ bead polymer.
The simulations were equilibrated for $10^6 \t$ before collecting data
over further $3\times 10^8\t$. We used periodic boundary conditions in $x$ and
$y$ directions with the lateral extent of the simulation box in these directions $W= 2L$.
In Fig.\ref{deGen1} we have plotted the three components of mean squared end-to-end distance
$\la X^2\ra'=\la X^2\ra-\l_\parallel$,
$\la Y^2\ra'=\la Y^2\ra-\l_\parallel$, and $\la Z^2\ra'=\la Z^2\ra+\l_\perp$ where
$\l_\parallel$ and $\l_\perp$ are offset polymer sizes which comes about for reasons discussed
in the following.

The mean squared
end-to-end distance in the confining direction $\la Z^2\ra'$
shrinks with increasing degree of confinement (decreasing $D$)
as $\g D^2$ (Fig.\ref{deGen1}).
Apart from an additive offset $\l_\perp$,
this is consistent with the leading order behavior predicted by our
MFT [Eq.~(\ref{eq:ro})] and de Gennes scaling.
Due to a finite non-zero range of repulsion $z_c$ coming from the
soft-wall confinement (not incorporated in the MFT as in the MFT we assumed
hard wall confinement for simplicity), simulated $\la Z^2\ra$ gets suppressed by
an extra amount $\l_\perp$.
This indicates that $\l_\perp$ should be a function of
$z_c$ which vanishes as $z_c \to 0$ (hard wall limit).
The fitting procedure in Fig.\ref{deGen1} gives $\l_\perp = 1.4 \pm 0.06$
which is numerically indistinguishable from $z_c^2=1.36$.
At larger $D$, we find a saturation of $\la Z^2\ra$ that obeys the functional
form (Fig.\ref{deGen1})
$$\chi(D)=\g D^2 - c_1 D^2 \exp(-c_2/D^2 + c_3/D^{1/2}).$$
Notice that this form of $\chi(D)$  is obtained from the
expression of $\la R_\perp^2\ra$ obtained from Eqs.~(\ref{eq:ro})
and (\ref{eq:lo}).

Fig.\ref{deGen1} clearly shows that  with increasing degree
of confinement, the components of mean squared end-to-end separation
in the unconfined directions expands
$\la X^2\ra' = \la Y^2\ra' = \mu/D^{1/2}$.
Leaving out the offset value $\l_\parallel$, this obeys de Gennes scaling.
Note that an offset like $\l_\parallel$ was expected from our MFT
[see the first line of Eq.~(\ref{eq:rp})].
Further, the excess compression of $\la Z^2\ra$ by $\l_\perp$
requires an extra bit of expansion in
the parallel components. $\l_\parallel$ contains this
contribution too.
At large enough plate-separations $\la X^2\ra$, $\la Y^2\ra$ saturate
to their bulk value.

The total mean squared end-to-end distance of the confined-polymer
$\la R_{\rm tot}^2 \ra = \la X^2\ra + \la Y^2 \ra + \la Z^2 \ra $
shows a non-monotonic dependence on $D$ (Fig.\ref{deGen2}).
$\la R_{\rm tot}^2 \ra$, starting from its bulk value at large $D$,
first reduces to reach a minimum and then expands as we reduce the
plate separation.
The initial decrease in size is mainly governed by the shrinkage
in $\la Z^2 \ra$.
At small $D$ values, the large increase of the polymer-size
in the unconfined directions $\la X^2 \ra$ and $\la Y^2 \ra$ takes over.
The cross-over between these
two different behaviors takes place at the minimum of $\la R_{\rm tot}^2 \ra$, a feature predicted by the
MFT [Eq.~(\ref{eq:dc})] as well as by de Gennes scaling (Sec.\ref{scaling}).
Fig.\ref{deGen2} shows that the scaling forms added with the offset parameters
in the three components of the mean squared end-to-end distance does capture
the non-monotonicity as well as the approximate position of the minimum in
$\la R_{\rm tot}^2 \ra$.

\section{Discussion and outlook}
\label{conc} In this paper we have presented a mean field approach to calculate the mean squared end-to-end distance
of a self-avoiding polymer confined within a parallel-plate geometry. The method allowed us to calculate all the components of
this quantity separately as a function of plate-separation. Up to the leading order calculation, we recovered Flory
scaling in the limit of large plate separations (bulk limit), and de Gennes scaling in the limit of small plate separations.
The next to leading order correction allows a smooth transition from de Gennes regime towards Flory regime. We
believe that higher order perturbation calculations using our MFT framework will further bridge the gap between the
two regimes. We noted that the non-monotonicity of the overall polymer size as a function of the increased degree of
confinement, as clearly captured by our MFT, was already inherent to de Gennes scaling. The numerical results from
Langevin dynamics simulations showed good agreement with our MFT predictions.

It is, however, hard to make quantitative comparisons between the theoretical calculations and the simulations for
quantities like actual prefactors of scaling laws. There are two reasons behind this. First, these non-universal
quantities depend on the detailed nature and strength of the interactions. Secondly, as was already discussed in
Ref.\cite{Edwards1979}, the effective values of the prefactors in the scaling forms depend on up to which order the
perturbative calculations are performed in the MFT.

Before we end the discussion, we would like to emphasize the simple fact that with decreasing plate-separations, the
polymer size in the confining direction  can only shrink, while it expands in the other unconfined directions. In a
recent experiment on DNA confined between two parallel plates~\cite{Bonthuis2008}, the average radius of gyration was
measured from the configurations projected onto a plane parallel to the confining glass plates.
Ref.~\cite{Bonthuis2008} showed that with reducing plate-separation $D$, this {\em projected} radius of gyration first shrank,
before expanding in accordance with de Gennes scaling $D^{-1/4}$. Similar non-monotonic features were also reported
in Ref.~\cite{Uemura2010}. This behavior should not be confused with the non-monotonicity in $\la R_{\rm tot}^2\ra$
(Fig.\ref{deGen2}) and is clearly not in agreement with our assertion mentioned above.
Ref.~\cite{Uemura2010} speculated that this non-monotonicity may be due to attractive interaction with the confining
walls, as the detailed nature of polymer-wall interaction is both hard to control and to determine in experiments.
It should be noted, however,  that the non-monotonicity in the radius of gyration calculated from the full 3d
configurations of a confined self-avoiding polymer, as observed in our Langevin dynamics simulations (data not
shown) and previous Monte-Carlo simulations~\cite{Vliet1990}, has a different origin. It is purely an entropic
effect caused by the hard confinement and, as we have shown in Sec.~\ref{scaling}, is already to be expected on the basis of de Gennes scaling.
Using our Langevin dynamics simulations, we have also evaluated the radius of gyration of the excluded volume polymer confined between two parallel plates from configurations projected onto a plane parallel to the plates, exactly in the same manner as described
in Ref.~\cite{Bonthuis2008}. These simulations showed a clear monotonic increase as $D^{-1/4}$ with reducing $D$
(data not shown). This is in contrast to the observed behavior in Refs.~\cite{Bonthuis2008,Uemura2010}, and in
agreement with our assertion that polymer size in the unconfined directions can only monotonically grow with
increasing degree of confinement.

Finally, the MFT approach used in this paper can be easily extended to other confining geometries, e.g., cubic or
cylindrical pores. The cylindrical confinement is a natural choice for the study of bacterial
chromosomes~\cite{Jun2006}. Within a biological cell, binding of proteins on DNA may enhance (or reduce) the
effective persistence length of DNA. Thus it would be interesting to examine the
effect of varying  bending stiffness on confined 
polymers. 
A similar MFT method may potentially also be used to describe tethered polymers. For this case one may need to
self-consistently determine the step-size of the effective Gaussian polymer as a function of distance from the
tethered end. Thus, this mean field approach has the potential to provide analytical access to many situations of
biological relevance.

\smallskip

\acknowledgments This work is part of the research program of the Stichting voor Fundamenteel Onderzoek der Materie
(FOM), which is supported by the Nederlandse organisatie voor Wetenschappelijk Onderzoek (NWO). DC is funded by
project $07$DNAA$02$ in the FOM-program $\#103$ ``DNA in action: Physics of the genome". We thank C. Dekker for
discussions on Ref.~\cite{Bonthuis2008}, and Nils Becker for a critical reading of the manuscript.

\smallskip
\appendix
\section{Calculating Green's function}
\label{ap-greens}
A separation of variable $G_0=f(L) \psi(\rp) \phi(z)$ in Eq.~(\ref{eq:gf0}) leads to,
\beq
\f{1}{f} \f{df}{dL} = \f{\lp}{6} \f{1}{\psi} \nabla_\parallel^2 \psi + \f{\lo}{6} \f{1}{\phi} \f{d^2 \phi}{dz^2} \equiv -E
\label{eq:sepa}
\eeq
where $E$ is an arbitrary constant.
%
Further, let us assume,
\bea
\left(\f{d^2}{dz^2} + k_z^2 \right) \phi(z) &=&0 \\
\label{eq:gperp}
\left( \nabla_\parallel^2 + \kp^2\right) \psi(\rp) &=&0.
\eea
The Dirichlet boundary condition $G_0(z=0,D)=0$ requires the solution,
$\phi_n(z)=\sqrt{2/D}\sin(k_z^n z)\, \sqrt{2/D} \sin(k_z^n z')$ with  $k_z^n=(n\pi/D)$,
$n$ being a positive integer. The solution in the $x$ and $y$
directions gives $\psi(\rp)=\exp(i \kp . \rp)$ with a continuum of $\kp$ modes in presence
of an open boundary condition.
Eq.~(\ref{eq:sepa}) leads to the identity,
\beq
E = \f{\lp}{6} \kp^2 + \f{\lo}{6} \left(\f{n\pi}{D} \right)^2.
\eeq
Thus the solution of the differential equation Eq.~(\ref{eq:gf0}) has the form
$\exp(-E L)\exp(i \kp.\rp) \sin(k_z^n z) \sin(k_z^n z')$. The
general solution is obtained by summing (integrating) over all possible
modes $n$  ($\kp$),
\begin{widetext}
\bea
G_0 &=& \f{2}{D} \int \f{d\kp}{(2\pi)^2} \sum_{n=1}^\infty e^{-(\lp/6)\kp^2 L - (\lo/6)(n\pi/D)^2 L} e^{i\kp . (\rp -\rp')}
 \sin\left( \f{ n\pi}{D} z \right) \sin\left( \f{ n\pi}{D} z' \right) \crcr
&=& \f{3}{2\pi \lp L} e^{-3(\rp -\rp')^2/2\lp L} \times \f{2}{D} \sum_{n=1}^\infty e^{- (\lo/6)(n\pi/D)^2 L}
\sin\left( \f{ n\pi}{D} z \right) \sin\left( \f{ n\pi}{D} z' \right) \equiv \Gp \times \Go.\nn\\
\eea
\end{widetext}


\section{Contributions from excluded volume interactions}
\label{ap:exvol}
In this appendix we calculate $\d_\parallel^b$ and $\d_\perp^b$.

The first order contributions due to the excluded volume term in the
Hamiltonian are of the form%
\begin{multline}
\left\langle A\left(  \mathbf{r}\left(  0\right)  ,\mathbf{r}\left(  L\right)
\right)  \be \Delta\mathcal{H}_{b}\right\rangle_0 =\\
\frac{1}{Z}\frac{1}{2}b\int d\mathbf{r}\left(  0\right)  \int
d\mathbf{r}\left(  L\right)  A\left(  \mathbf{r}\left(  0\right)
,\mathbf{r}\left(  L\right)  \right)  \int_{0}^{L}ds\int_{0}^{L}ds^{\prime}\\
\int_{\mathbf{r}\left(  0\right)  }^{\mathbf{r}\left(  L\right)  }%
\mathcal{D}[\mathbf{r}(s)]\delta\left(  \mathbf{r}(s)-\mathbf{r}(s^{\prime
})\right)  \exp\left\{  {-\beta\mathcal{H}_0[\mathbf{r}(s)]}\right\} \nonumber\\
\equiv\frac{1}{Z}\frac{1}{2}b\int d\mathbf{r}\left(  0\right)  \int
d\mathbf{r}\left(  L\right)  A\left(  \mathbf{r}\left(  0\right),\mathbf{r}\left(  L\right)  \right) \\ \int_{0}^{L}ds\int_{0}^{L}ds^{\prime
}E\left(  s,s^{\prime}|\mathbf{r}\left(  0\right)  ,\mathbf{r}\left(
L\right)  \right), \nonumber
\end{multline}
where the integration over end-points are performed for all possible $\mathbf{r}(0)$ and $\mathbf{r}(L)$.
We consider the evaluation of the kernel $E\left(  s,s^{\prime}|\mathbf{r}%
\left(  0\right)  ,\mathbf{r}\left(  L\right)  \right)  $ of the inner
integration, assuming for the moment that $s>s^{\prime}.$ The
Chapman-Kolmogorov property
\begin{equation}
G\left(  \mathbf{r,}L|\mathbf{r}^{\prime},0\right)  =\int d\mathbf{r}%
^{\prime\prime}G\left(  \mathbf{r,}L|\mathbf{r}^{\prime\prime},L-s\right)\,
G\left(  \mathbf{r}^{\prime\prime}\mathbf{,}s|\mathbf{r}^{\prime},0\right)
\label{eq:CK}%
\end{equation}
then allows us to write%
\bea
E(s,s' \mid \rv (0), \rv (L)) &=& \int d\rv' G_0(\rv(L),L \mid \rv', s) \crcr
&& G_0(\rv',s\mid \rv',s')\, G_0(\rv',s'\mid \rv(0),0).\nn\\
\eea
We also use the fact that since the Green's function factorizes as $G_0=G_\parallel G_\perp$, the separation of variable
works for $E=\Ep \times \Eo $ as well.

As the longitudinal part is
unaffected by the constraints,  it is the simplest case. Indeed, as%
\begin{equation}
G_{\Vert}\left(  \mathbf{r}^{\prime}\mathbf{,}s|\mathbf{r}^{\prime},s^{\prime
}\right)  =\frac{3}{2\pi l_{\Vert}\left(  s-s^{\prime}\right)  }%
\end{equation}
\textit{does not} depend on the intermediate position $\mathbf{r}_{\Vert
}^{\prime}$ we immediately find that%
\begin{multline}
\Ep \\
= \f{3}{2\pi\lp (s-s')} \Gp(\rp(L), L-(s-s') \mid \rp(0), 0)\\
\end{multline}
A similar simplification does not hold for the transverse component as%
\begin{equation}
G_{\bot}\left(  r_{\perp}^{\prime}\mathbf{,}s|r_{\perp}^{\prime},s^{\prime
}\right)  =\frac{2}{D}\sum_{n=1}^{\infty}\sin^{2}\left(  \frac{n\pi}%
{D}r_{\perp}^{\prime}\right)  e^{-\k n^2\left(  s-s^{\prime}\right)  }%
\end{equation}
(with $\k = (\lo/6)(\pi/D)^2$)
\textit{does} depend on $r_{\perp}^{\prime}.$ We therefore find that%
\begin{multline}
 \Eo \\
=\f{1}{D^2} \sum_{n_L,n',n_0=1}^\infty e^{-{\k} L n_L^2} e^{-{\k} s (n'^2-n_L^2)} e^{-{\k} s' (n_0^2-n'^2)} \\
 \times [2\d_{n_0,n_L} + \d_{n',(n_0+nL)/2} - \d_{n',(n_0-nL)/2} -\d_{n',(nL-n_0)/2} ]\\
  \times \sin\left(\f{n_L\pi}{D} \ro(L)\right) \sin \left(\f{n_0\pi}{D} \ro(0)\right),
\end{multline}

Again, we can write
\bea
\la R_\parallel^2 \be \D {\cal H}_b\ra_0 &=& Z^{-1} \f{1}{2} b \int_0^L ds \int_0^L ds' \int d\rv (0) \int d\rv (L) \crcr
&& ~[\rp (L) - \rp (0)]^2~
E(s,s' \mid \rv (0), \rv (L)) \crcr
&=& Z^{-1} \f{1}{2} b \int_0^L ds \int_0^L ds' \crcr
&& \left[\int d\rp (0) \int d\rp (L) ~[\rp (L) - \rp (0)]^2~  E_\parallel \right]\crcr
&& \left[\int d\ro (0) \int d\ro (L)  E_\perp \right]\crcr
&\equiv& Z^{-1} \f{1}{2} b \int_0^L ds \int_0^L ds' \ipt \ioo.\nn\\
\eea
Similarly, 
\bea
\la R_\perp^2 \be \D {\cal H}_b\ra_0 &=&  Z^{-1} \f{1}{2} b \int_0^L ds \int_0^L ds' \ipo \iot.\nn\\
\\
\la \be \D {\cal H}_b\ra_0 &=& Z^{-1} \f{1}{2} b \int_0^L ds \int_0^L ds' \ipo \ioo. \nn\\
\eea
In the above, we used the  definitions
\bea
\ipo &=& \int d\rp(0) \int d\rp(L) ~\Ep \nn\\
\label{eq:ip0}
\\
\ipt &=& \int d\rp(0) \int d\rp(L) ~\Ep \nn\\
&\times& (\rp(L)-\rp(0))^2 
  \label{eq:ip2}\\
\ioo &=& \int d\ro(0) \int d\ro(L) ~\Eo \nn\\
\label{eq:io0}\\
\iot &=& \int d\ro(0) \int d\ro(L) ~\Eo \nn\\
&\times& (\ro(L)-\ro(0))^2.
\label{eq:io2}
\eea

We can now evaluate the integrals over the end points. The longitudinal
ones are easier to compute,
\bea
\ipo &=& W^2 \f{3}{2\pi \lp (s-s')} \crcr
\ipt &=& W^2 \f{3}{2\pi \lp (s-s')} \times \f{2}{3} \lp (L-(s-s')).
\label{eq:ip02}\nn\\
\eea

The only $\ro(0)$ and $\ro(L)$ dependent term present in $\Eo$ is
$q(\ro(L),\ro(0))=\sin (n_L\pi \ro(L)/D)\times \sin (n_0\pi \ro(0)/D)$. To evaluate the integration in $\ioo$ we use
the identity
\beq
\int_0^D d\ro \sin\left( \f{n\pi}{D} \ro \right) = \f{D}{n\pi}[1-(-1)^n].
\eeq

Thus,
\bea
\ioo &=&  \sum_{n_L,n',n_0=1}^\infty e^{-{\k} L n_L^2} e^{-{\k} s (n'^2-n_L^2)} e^{-{\k} s' (n_0^2-n'^2)} \nn\\
         &\times& [2\d_{n_0,n_L} + \d_{n',(n_0+n_L)/2} \nn\\ &-& \d_{n',(n_0-n_L)/2} -\d_{n',(n_L-n_0)/2} ] J_0(n_L,n_0)\nn\\
\eea
where
\beq
J_0(n_L,n_0)=\f{1}{n_0 n_L\pi^2} [1-(-1)^{n_L}] [1-(-1)^{n_0}].
\eeq
To calculate $\iot$ one requires to use the integration of sine function with powers, such as,
$\int_0^D dr\, \sin(n\pi r/D)$, $\int_0^D dr\, r\, \sin(n\pi r/D)$, and $\int_0^D dr\, r^2\, \sin(n\pi r/D)$.
This gives us,
\bea
\iot &=& D^2 \sum_{n_L,n',n_0=1}^\infty e^{-{\k} L n_L^2} e^{-{\k} s (n'^2-n_L^2)} e^{-{\k} s' (n_0^2-n'^2)} \nn\\
         &\times& [2\d_{n_0,n_L} + \d_{n',(n_0+nL)/2} \nn\\ &-& \d_{n',(n_0-nL)/2}  -\d_{n',(nL-n_0)/2} ]\nn\\
          &\times&J(n_L,n_0),
\eea
where,
\bea
J(n_L,n_0) &=&- \f{1}{n_L^3 n_0\pi^4} [(-1)^{n_L} n_L^2\pi^2 +2(1-(-1)^{n_L})]\nn \\
                     &\times& [1-(-1)^{n_0}] - \f{2}{n_L n_0 \pi^2} (-1)^{n_0}(-1)^{n_L}\nn\\
                     &-& \f{1}{n_L n_0^3\pi^4} [(-1)^{n_0} n_0^2\pi^2 +2(1-(-1)^{n_0})] \nn\\
                     &\times& [1-(-1)^{n_L}].
\eea

We used $s>s'$ in the above.
Now, using the identity
\beq
\f{1}{2} \int_0^L ds \int_0^L ds' = \int_0^L ds' \int_{s'}^L ds
\eeq
we have
\bea
\d_\parallel^b &=& Z^{-1} b \int_0^L ds' \int_{s'}^L ds~ \ioo [\ipt - \la R_\parallel^2\ra_0 \ipo] \nn\\
\label{eq:dpb}
\\
\d_\perp^b &=& Z^{-1} b  \int_0^L ds' \int_{s'}^L ds~ \ipo [\iot - \la R_\perp^2\ra_0 \ioo]. \nn\\
\label{eq:dob}
\eea
We now have all the ingredients to calculate the perturbation corrections.

\subsection{Parallel Component $\d_\parallel^b$}
We have,
\begin{multline}
[\ipt - \la R_\parallel^2\ra_0 \ipo]\\
=  \f{3 W^2}{2\pi\lp (s-s')}
\times  \left(\f{2}{3}\lp(L-(s-s')) - \f{2}{3}\lp L \right)
= - \f{W^2}{\pi}.\\
\end{multline}
Thus in evaluating the integral in $\d_\parallel^b$ (Eq.~\ref{eq:dpb}), we need to do the
following integration,
\beq
M =  e^{-{\k} L n_L^2} \int_0^L ds' \int_{s'}^L ds~ e^{-{\k} s (n'^2-n_L^2)} e^{-{\k} s' (n_0^2-n'^2)}
\eeq
where  the only function of $(s,s')$ in the integrand comes from $\ioo$.

Thus, we obtain,
\bea
\d_\parallel^b &=& -\f{b}{Z}W^2 \f{1}{\pi} \sum_{n_L,n',n_0=1}^\infty M(n_L,n',n_0;{\k}) \nn\\
           &\times& [2\d_{n_0,n_L} + \d_{n',(n_0+nL)/2} \crcr
           &&  - \d_{n',(n_0-nL)/2}  -\d_{n',(nL-n_0)/2} ]\, J_0(n_L, n_0), \nn
\eea
i.e., Eq.~\ref{dpll_b}.

\subsection{Perpendicular Component $\d_\perp^b$}
%
%
The only function of $(s,s')$ in Eq.~\ref{eq:dob} is
$\exp[-{\k} s (n'^2-n_L^2)]\exp[-{\k} s' (n_0^2-n'^2)]$ due to the function
$[\iot-\la R_\perp^2\ra_0 \ioo]$  and $1/(s-s')$ due to
$\ipo$. Thus we have to evaluate the integration,
\beq
K=e^{-{\k} L n_L^2} \int_0^L ds' \int_{s'}^L ds \f{e^{-{\k} s (n'^2-n_L^2)} e^{-{\k} s' (n_0^2-n'^2)}}{(s-s')}.
\label{eq:K}
\eeq
The  term by term calculation of this quantity for each ($n_0,n',n_l$) encounters a pole at $s=s'$.
To avoid the pole, the integration to evaluate Eq.~(\ref{eq:K}) can be rewritten as,
\bea
K(n_L,n',n_0;{\k}) &=& e^{-{\k} L n_L^2}\, \lim_{a\to 0}\, \int_0^L ds' \int_{s'}^L ds \crcr
&& \left[\f{e^{-{\k} s (n'^2-n_L^2)} e^{-{\k} s' (n_0^2-n'^2)}}{(s-s')+ a}\right]. \nn\\
\eea
Note that this is equivalent to replacing the delta-function
overlap-interaction by its Gaussian representation but keeping a non-zero
variance $a^2$.
%
Thus we obtain the expression (Eq.~\ref{dporth}),
\bea
\d_\perp^b &=& \f{b}{Z} \,W^2\, \f{3}{2\pi\lp} \sum_{n_L,n',n_0=1}^\infty D^2\,  K(n_L,n',n_0;{\k}) \nn\\
           &\times& [2\d_{n_0,n_L} + \d_{n',(n_0+nL)/2} \crcr
            &-& \d_{n',(n_0-nL)/2}  -\d_{n',(nL-n_0)/2} ] \crcr
           &\times&  \left[ J(n_L,n_0) - \f{\la R_\perp^2 \ra_0}{D^2} J_0(n_L,n_0) \right].\nn
\eea


\section{Bulk limit of $D\to \infty$}
\label{apfree}
In this appendix we calculate $\d_\parallel^b$, $\d_\perp^b$ and
$\d_\perp^\perp$ in the bulk limit of $D\to \infty$.

It is clear that, since in this limit $\Go$ itself is independent of $D$,
$\d_\parallel^b$ and $\d_\perp^b$ are also independent of $D$.
Using the continuum approach at $D\to \infty$,
\bea
E_\perp &=& \sqrt{\f{3}{2\pi\lo (s-s')}} \Go(z(L), L-(s-s') \mid z(0), 0) \crcr
&=& \sqrt{\f{3}{2\pi\lo (s-s')}} \times \sqrt{\f{3}{2\pi\lo (L-(s-s'))}}\crcr
 &\times& \exp \left[-\f{3}{2}\f{(z(L)-z(0))^2}{\lo (L-(s-s'))}\right].
\eea
Then using Eq.~\ref{eq:io0} we get
\bea
\ioo &=& \int dz(0) \int dz(L) \Eo\crcr
&=& D \sqrt{\f{3}{2\pi \lo (s-s')}}.
\eea
Remember that, now, the orthogonal component of partition function has also got
redefined to $\Zo = D$.
Then using Eq.~\ref{eq:dpb},
\bea
\d_\parallel^b &=& Z^{-1} b \int_0^L ds' \int_{s'}^L ds~ \ioo [\ipt - \la R_\parallel^2\ra_0 \ipo] \crcr
&=& (W^2 D)^{-1} b (-W^2/\pi) \crcr
&\times& D \sqrt{\f{3}{2\pi\lo}} \int_0^L ds' \int_{s'}^L ds~ \sqrt{\f{1}{(s-s')}}\crcr
&=& -\f{b}{\pi}\times \sqrt{\f{3}{2\pi\lo}} \times \f{4}{3} L^{3/2};
\eea
where we also used Eqs.~\ref{eq:ip02} and \ref{eq:Rpareval}.
Using Eq.~\ref{eq:io2} we find 
\bea
\iot &=& \int_{-\infty}^\infty dz(0) \int_{-\infty}^\infty dz(L) \sqrt{\f{3}{2\pi\lo (s-s')}}\crcr
&\times& \sqrt{\f{3}{2\pi\lo (L-(s-s'))}} e^{-\f{3(z(L)-z(0))^2)}{2\lo (L-(s-s'))}} \crcr
&=& \sqrt{\f{3}{2\pi\lo (s-s')}} D \f{\lo (L-(s-s'))}{3} 
\eea
and using Eq.~\ref{eq:Rprpeval} we get
\bea
\la R_\perp^2\ra_0 &=& \f{\int_{-\infty}^\infty dz \int_{-\infty}^\infty dz' (z-z')^2  e^{-\f{3(z(L)-z(0))^2)}{2\lo L}}} {\int_{-\infty}^\infty dz \int_{-\infty}^\infty dz'  e^{-\f{3(z(L)-z(0))^2)}{2\lo L}}} \crcr
&=& \f{\lo L}{3}.
\label{rpB}
\eea
Therefore,
\beq
\iot-\la R_\perp^2\ra_0 \ioo = -D \f{\lo}{6\pi} (s-s')^{1/2},
\eeq
and
\begin{align}
&\ipo [\iot-\la R_\perp^2\ra_0 \ioo] \nn\\
&= -W^2 \f{3}{2\pi \lp (s-s')} D \sqrt{\f{\lo}{6\pi}} (s-s')^{1/2} \nn\\
&-W^2 \f{D \lo^{1/2}}{\lp} (s-s')^{-1/2}.
\end{align}
Since in this limit $Z=W^2 D$, using Eq.~\ref{eq:dob} we find,
\bea
\d_\perp^b &=& Z^{-1} b  \int_0^L ds' \int_{s'}^L ds~ \ipo \nn\\
&\times& [\iot - \la R_\perp^2\ra_0 \ioo] \crcr
&\simeq & -\f{b\lo^{1/2}}{\lp} L^{3/2}
\eea
Similarly, using Eq.~\ref{rpB} in Eq.~\ref{eq:doo}, we obtain
\bea
\d_\perp^\perp = - \f{3}{2} \left(\f{1}{l} -\f{1}{l_\perp}\right) \f{\partial}{\partial \a_\perp}\la R_\perp^2\ra_0
= \f{1}{3}\left(\f{1}{l} -\f{1}{l_\perp}\right) L \lo^2.\nn\\
\eea

\section{2d limit of confined system}
\label{ap:2d}
In pure 2d, the perturbative contribution due to the
inter-segment interactions takes the form,
\beq
\d_{2d}^b = Z_{2d}^{-1} b_0 \int_0^L ds'\int_{s'}^L ds [\ipt-\la R_\parallel^2\ra_0 \ipo] \nn
\eeq
where $Z_{2d}=W^2$. Thus,
\beq
\d_{2d}^b 
= -\f{b_0}{2\pi}L^2 \nn
\eeq
where $b_0$ plays the role of interaction strength.
Thus, $\d_\parallel =\d_\parallel^\parallel+\d_\parallel^b=0$
implies
\bea
\f{L l_{2d}^2 }{l} &\sim& \f{b_0}{2\pi}L^2 \crcr
\mbox{or,~~} l_{2d} &\sim& (b_0 l)^{1/2} L^{1/2} \nn
\eea
which leads to the scaling form,
\beq
{\la R_{2d}^2\ra} \sim L l_{2d} \sim (b_0 l)^{1/2} L^{3/2}.
\eeq

We expect that the polymer would behave like a pure 2d polymer in the
limit of extremely small plate-separation $D$. This limit is achieved when
$\la R_\parallel^2\ra = \la R_{2d}^2\ra$.
We have seen that the strength of the interaction $b$ in 3d has the
dimension of length, whereas the same strength $b_0$ is dimension-less
in 2d. This shows that the only
intrinsic microscopic length scale in the system is segment-length  $l$. Thus
we need to express $b=b_0 l$ to search for the pure 2d limit of confined systems.
With this substitution,
\beq
\la R_\parallel^2\ra \sim (b_0 l)^{1/2} L^{3/2} \left(\f{l}{D}\right)^{1/2}.
\eeq
Therefore, $\la R_\parallel^2\ra$ equates $\la R_{2d}^2\ra$ when $D\sim l$, i.e., when the
plate-separation $D$ becomes as small as the polymer segment-length  $l$.

\bibliographystyle{apsrev4-1}

%
\end{document}